\documentclass[12pt]{article}
\begin{document}
\title{{\bf Possible Anthropic Support for a Decaying Universe:\\
A Cosmic Doomsday Argument}
\thanks{Alberta-Thy-12-09, arXiv:0907.4153}}
\author{
Don N. Page
\thanks{Internet address:
don@phys.ualberta.ca}
\\
Theoretical Physics Institute\\
Department of Physics, University of Alberta\\
Room 238 CEB, 11322 -- 89 Avenue\\
Edmonton, Alberta, Canada T6G 2G7
}
\date{(2009 July 24)}

\maketitle
\large
\begin{abstract}
\baselineskip 18 pt

I have suggested that one possible solution of the Boltzmann brain problem is that the universe is decaying at an astronomical rate, making it likely to decay within 20 billion years.  A problem with this suggestion is that it seems to require unnatural fine tuning in the decay mechanism that would not be explained anthropically.  Here it is pointed out that if a spacetime version of volume averaging were used in the cosmological measure problem, this would give anthropic support for an impending cosmic doomsday.

\end{abstract}
\normalsize

\baselineskip 13.9 pt

\newpage

\section*{Introduction}

One of the main challenges of theoretical cosmology today is the {\it measure problem} \cite{LLM,BLL94,Vil95a,GL,LLM2,Vil95b,WV,LM,Page97,Vil98,VVW,Guth00,GV,EKS,SEK, AT,Teg,Aguirre,Ellis,Page-in-Carr,GSVW,ELM,BFL,Bou06,CDGKL,BFY06,GT, Vil07,AGJ1, Win06,AGJ2,BHKP,Guth07,BY,LWa,ADNTV,Linde08a,Linde08b,LWb,CSS, Haw,GV07,HHH, Win08a,SGSV,BFYb,DGLNSV,GV09,Win08b,Wang,Win08c,McI,LVW, HaHe,SH,Albrecht09}, how to get normalized probabilities for observations in a universe that might simply be or have inflated to become arbitrarily large and have arbitrarily many observers and observations.  An acute aspect of the measure problem is the question of how to show that observations are not dominated by disordered Boltzmann brain observations, which would seem to swamp ordinary observers if the universe lasts too long 
\cite{DKS,Albrecht04,AS,Page05,YY,Page06a,BF,Page06b,Bou05,Linde06,Page06c, Vil06,Page06d,Vanchurin,Banks,Carlip,HS,GM,Giddings,typdef,LWc,DP07,Bou08, BFYa,Gott,typder,FL}.  I have suggested \cite{Page05,Page06a,Page06c,Page06d} that one possible solution of the Boltzmann brain problem would be for the universe to be decaying at an astronomical rate, so that it would not be expected to last more than 20 billion years from now.  However, proposed mechanisms for this decay \cite{Page05,Page06a,ADSV} appears to require a high degree of fine tuning that seemed inexplicable.  In particular, past methods of conditioning upon the existence of observers do not seem to help to explain this fine tuning, so it has not appeared to have an anthropic explanation.

Here it is suggested that if the volume weighting that I proposed in \cite{cmwvw} (see also \cite{insuff,brd,BA} for further discussion and motivation) is applied to the spacetime 4-volume of complete histories of the universe rather than to the spatial 3-volumes of hypersurfaces of constant time as I have done previously, then the probabilities of observations may be likely to be dominated by histories in which the lifetime is not enormously greater than that within our observed past.  This would then give an anthropic argument in favor of universes that last only a finite time, thus making more viable my previous proposed solution to the Boltzmann brain problem.  Furthermore, if the parameter that gives the astronomical decay rate is observationally or experimentally accessible (e.g., if it were related to the cosmic parameter $w$ \cite{Page05,Page06a} or a measurable property of the Higgs particle \cite{ADSV}), this prediction of cosmic doomsday could be testable, so that in principle we could gain more direct evidence that our universe is facing decay.

\section{Previous arguments for and against astronomical decay rates}

Let me briefly summarize my arguments \cite{Page05,Page06a,Page06c,Page06d} for and against a rapid or astronomical decay rate of the universe, where by rapid or astronomical I mean on the scale of the present age, so that the expected future lifetime would be of the order of $10^{10}$ years or $10^{61}$ Planck times, rather than having an exponentially large number in the exponent, such as one would get from the quantum recurrence time $e^{10^{122}}$ for our apparently asymptotically de Sitter spacetime.

The basic argument for rapid decay is that it would solve the Boltzmann brain (BB) problem.  Without such decay, one can state the BB problem as follows \cite{Page05,Page06a,Page06c,Page06d}:

Consider a finite comoving volume of an ever-expanding universe.  Presumably ordinary observers (OOs) depend upon free energy from stars and so die out when stars do, giving a finite total number of OOs per comoving volume.  However, de Sitter thermal fluctuations, or even just vacuum quantum fluctuations, presumably give some extremely tiny but positive probability per 4-volume for BBs to fluctuate into existence and make observations.  If the universe lasts for an infinite time, this tiny rate multiplied by the infinite 4-volume will lead to an infinite number of BBs per comoving volume, utterly swamping the finite number of OOs per comoving volume.

This would apparently imply that statistically one would expect any random observation to be a BB observation.  However, BB observations are not expected to have the order we observe for our observations, so our observations are not consistent with being a random observation if nearly all observations are BB observations.  Therefore, making the plausible assumption that our observations are randomly chosen from the set of all observations, statistically our ordered observations rule out any scenario in which BBs swamp OOs, as would seemingly be the case in a fixed comoving volume if the universe lasts forever and has only a finite time during which OOs can exist.

To get a lower limit on the decay rate in the case that it is by the quantum nucleation of a bubble, at annihilation probability $A$ per 4-volume, that then expands at nearly the speed of light \cite{Page06a,Page06d}, one calculates that the minimum $A$ to prevent the expected future 4-volume of a comoving volume from being infinite is $A_\mathrm{min} = 9H_\Lambda^4/(4\pi) = \Lambda^2/(4\pi) \approx (18\pm 2\ \mathrm{Gyr})^{-4}$, where $H_\Lambda = H_0\sqrt{\Omega_\Lambda} \approx (16\pm 2\ \mathrm{Gyr})^{-1}$ is the asymptotic Hubble rate in the future de Sitter era if the presently observed dark energy is indeed a cosmological constant ($w\equiv p/\rho = -1$).  This would then imply that the future half-life of our universe would be less than 19 Gyr.

There are various objections to the assumptions in this argument (e.g., especially in \cite{Vil06,Smolin,BS,Banks,Carlip,HS,Gott,MA}, but also more indirectly in papers cited later in this paragraph).  For example, in the picture of eternal inflation, new bubbles continue to form arbitrarily far into the future and produce more OOs, so there is not only a finite time during which OOs can exist.  Then one has to choose some regularization of the infinite numbers per comoving volume of both OOs and BBs to make the OOs dominate, but proposals for doing this have been made \cite{BF,Bou05,Linde06,Vanchurin,GM,LWc,BFYa,Win08a,Win08b, Wang,FL,BFYb,DGLNSV, GV09,Win08c,McI,LVW,SH,Albrecht09}.  However, in this paper I do not wish to get into these difficult issues of eternal inflation but rather focus on a problem raised within the context of my proposed solution with its assumed rapid decay of the universe.

The problem within my proposal (mentioned within two of my papers on it \cite{Page06a,Page06d}) is that if the annihilation or decay rate $A$ is set by some microphysical parameter, then it seems that such a parameter must be fine tuned in a way that is not explained, even anthropically.  For example, I considered the possibility \cite{CDGKL} that the decay rate is set by the gravitino mass, which I then found \cite{Page06a,Page06d} had to be set to within a relative uncertainty of less than about 0.006.

Nima Arkani-Hamed, Sergei Dubovsky, Leonardo Senatore, and Giovanni Villadoro \cite{ADSV} have pointed out an interesting alternative decay mechanism for realizing the cosmic decay I predicted that would involve parameters of Higgs physics that in principle could be measured in the foreseeable future (unlike the nearly $10^{18}$ GeV gravitino mass needed for it to give suitable decay rates).  However, again there would have to be fine tuning, as the Higgs mass would need to be in a certain 0.1 GeV window depending upon the top mass and the strong interaction coupling constant.  (See also \cite{EEGHR} for a further discussion of this decay mechanism.)

One might note that other parameters of physics seem to be fine tuned, most dramatically the cosmological constant to zero within roughly one part in $10^{122}$ of the Planck value.  However, it is plausible that most of these other fine tunings can be explained anthropically.  One could presumably explain an upper limit on the decay rate $A$ anthropically from the requirement that the universe last long enough to produce ordinary observers in solar systems that take a long time to form, say \cite{Page06a,Page06d} $A_\mathrm{max} \sim 1000 A_\mathrm{min}$ in terms of the value of $A_\mathrm{min}$ given above, but before the ideas to be presented below, I could see no similar anthropic reason to explain a lower limit on $A$, since ordinary observers presumably could exist no matter low the decay rate is.

Furthermore, even if it turned out that the decay rate in our part of the multiverse is within the range $A_\mathrm{min} \leq A \leq A_\mathrm{max}$, it would still leave it mysterious why one does not have $A < A_\mathrm{min}$ in some other part of the multiverse that also allows Boltzmann brains \cite{Page06a,Page06d}.  If indeed there are other parts with $A < A_\mathrm{min}$ (with $A_\mathrm{min}$ equal to the value of $\Lambda^2/(4\pi)$ in that part of the multiverse), then it would seem that that part would have an infinite number of Boltzmann brains that would swamp the ordinary observers in our part.  In other words, we need to prevent BBs from dominating OOs over the entire multiverse, and not just in our part, so just having $A_\mathrm{min} \leq A$ in our part alone would not be sufficient.

\section{Rapid universe decay supported by 4-volume averaging}

The new idea of this paper is that if the volume averaging I proposed and discussed in \cite{cmwvw,insuff,brd,BA} is taken to be 4-volume averaging rather than the 3-volume averaging originally advocated there, this might select mainly histories of the universe that do not last long enough for Boltzmann brains to become a problem.

Traditionally many of us have believed that after weighting each component of the quantum state by absolute square of its amplitude, for calculating the probabilities of observations, one should also weight by the number of observers in that component.  For example, this is done explicitly in \cite{Page97,Haw,HHH} and at least implicitly by several proponents of eternal inflation \cite{Vet,Let,Set,Get}.  For a constant density of observers, this effectively weights the quantum components by their volumes as well as by the squares of their amplitudes, so I have called this volume weighting \cite{cmwvw,insuff,brd,BA}.

However, for an eternally expanding universe in which Boltzmann brains form at a positive (even if extremely tiny) rate per 4-volume, volume weighting leads to domination by the components of the quantum state (say eigenstates of the spatial geometry, as in canonical quantum gravity) with arbitrarily large volume and hence arbitrarily large numbers of observers in the form of Boltzmann brains.  Therefore, I proposed what I called volume averaging \cite{cmwvw,insuff,brd,BA}, weighting each quantum component not by the total number of observers, but by the spatial density of observers.  (This would be what one would get if one imagined choosing a random observation by sampling one random region of space, of fixed size, for each component of the quantum state, rather than sampling in each component a different number of regions proportional to the volume, which would give volume weighting.)  This volume averaging then would count components of the quantum state at very late times and with very large volumes and numbers of Boltzmann brains as each having very low weight in comparison with components corresponding the present much smaller size in which the spatial density of observers is much higher from the existence of ordinary observers.

Nevertheless, there remained a problem \cite{cmwvw} in that although the contribution from each component of the quantum state at huge volume (with negligible ordinary observers there) is very small in comparison with those of 3-geometries corresponding to the present universe with its much higher density of observers (almost entirely ordinary observers), if the universe expands forever, there will be infinitely more components of large size than those corresponding to the present universe.  The ratio depends on how one does the integral over the time in counting the different components that in canonical quantum gravity correspond to different times, but if one takes the most natural prescription \cite{HP}, one gets a weighting proportional to the proper time in the semiclassical limit, so then with an infinite amount of proper time to integrate over for an infinitely expanding universe, the total contribution of all of the infinitely many times of very large components with the Boltzmann brains would dominate over the finite times of the small components with ordinary observers.  Thus the infinite integral over time would cause the Boltzmann brains to dominate in the final result and make our observations of order statistically extremely improbable.

The new idea here is to do the volume averaging over the entire 4-volume of each history of the universe given by the quantum state, so that one weights each component of the state by the number of observers per 4-volume instead of per 3-volume on components of fixed 3-geometry.  In canonical quantum gravity the quantum state may be represented by a wavefunctional over 3-geometries and matter field configurations on each 3-geometry, so I do have my doubts whether it makes sense to interpret a quantum state of the universe as giving amplitudes or probabilities for 4-geometries (histories of 3-geometries) rather than just for 3-geometries themselves.  Thus I do have the fear that what I am proposing here may be nonsense, but since it is rather unclear what to do in quantum gravity, let us here suppose that it makes sense for a quantum state to give a quantum probability distribution over 4-geometries.

Indeed such a proposal of getting probabilities for 4-geometries (histories) rather than for 3-geometries (configurations) has been made within the decoherent-histories approach
\cite{Griffiths,Omnes,GellMann-Hartle,Hartle,Albrecht,Halliwell,Isham, Goldstein-Page,Isham-Linden,Diosi,Dowker-Kent,HLM,Kent,Kent-McElwaine, Hartle-Marolf,Griffiths-Hartle,Brun-Hartle,Markopoulou,Savvidou,Halliwell-Thorwart, Craig-Hartle,Halliwell-Wallden}, 
within which one can define probabilities for coarse-grained histories within a decoherent set.  This approach is plagued by the ambiguity of which decoherent set to take (see particularly \cite{Albrecht,Dowker-Kent,Kent,Kent-McElwaine}), since generally there are large families of them, so that I have admittedly been sceptical that this is the correct approach to take, but here for the sake of argument let us suppose that one can indeed use the quantum state of the universe to get a quantum probability for each member of a set of histories or 4-geometries.

So far, this probability distribution does not directly give the probability of observations, since one needs to know the distribution of observations within each history and how to weight them \cite{SH}, which is the classical analogue (within the set of histories interpreted as being classical 4-geometries) of the failure of Born's rule in cosmology \cite{insuff, brd,BA}.  This weighting of course is a weighting beyond the quantum weighting assigned to each history by the quantum state, which only gives a probability for each history and not the probability for each observation within the history.  (E.g., within a single history, one might want to weight an observation that occurs twice by twice the weight of an observation that occurs just once, and indeed I shall do that here.)

Here I am proposing that if one can indeed, from the quantum state of the universe, get a probability distribution over 4-geometries (including the internal spacetime history of the non-gravitational fields within each 4-geometry, though for short I shall sometimes refer to the entire history of all the fields as a 4-geometry), for observational probabilities (the probabilities of observational results, called observations for short), one should further weight by the density of observations on a 4-volume basis.  That is, take the probability of each possible observation as being proportional to the sum, over all histories, of the quantum probability of that history multiplied by the number of times the particular observation occurs within that history and divided by the total 4-volume of that history.

In the end one wants normalized probabilities of observations $O_j$ in a theory $T_i$ (which should in this scheme specify the quantum state, the rule for getting the probabilities of the histories from that quantum state [say by some particular choice of a set of decohering histories], and the rule for getting observational probabilities from that distribution of histories [here assumed to be by 4-volume averaging],
\begin{equation}
P_j(i) \equiv P(O_j|T_i)\ \mathrm{with}\  \sum_j P_j(i) = 1.
\label{prob}
\end{equation}
For this, one needs to divide each unnormalized observational probability $p_j(i)$ by the sum of all the unnormalized probabilities.

Thus the formula for observational probabilities in 4-volume averaging would be
\begin{equation}
P_j(i) = \frac{p_j(i)}{\sum_{k>0}p_k},
\label{normalized-probabilities}
\end{equation}
where each unnormalized observational probability would be
\begin{equation}
p_j(i) = \sum_h \frac{P(h|i)N(j|h)}{V_4(h)}.
\label{unnormalized-probabilities}
\end{equation}
Here $P(h|i)$ is the quantum probability that the quantum state given by the theory $T_i$ results in the history $h$, $N(j|h)$ is the number of times the particular observation $O_j$ occurs within the history $h$, and $V_4(h)$ is the 4-volume of the history $h$.  For ease of exposition, I am taking the set of histories $h$ to be discrete, so that one simply has a sum over them of the quantum probability of the history multiplied by the 4-volume density of the observation $O_j$, but of course one could take the set of histories to be continuous and replace the sum over $h$ by an integral.

Now suppose that one considers two possible histories of the universe, $h=1$ in which the universe expands for a very, very long time as an asymptotically de Sitter spacetime, and $h=2$ in which it is like our present universe up to the present, say 14 Gyr after the beginning, and then totally decays after twice this time, with a total lifetime of, say, 28 Gyr.  In history 1, in which the universe lasts long enough for Boltzmann brains in the late phase to dominate over the ordinary observers that last for only a finite time, $N(j|h)/V_4(h)$ is approximately given by the 4-volume density of Boltzmann brains, which is very roughly the exponential of the negative of the action to produce one, perhaps $e^{-10^{42}}$ \cite{Page06b} (maybe the right order of magnitude for the exponent of the exponent).  On the other hand, in history 2 most observers are presumably ordinary observers, and the 4-volume density would be much higher, perhaps something of the crude order of $e^{-536}$ in Planck units (maybe the right order of magnitude for the exponent itself) if one took, say, $10^{11}$ human observers in a 4-volume corresponding to a hypercube of length 14 billion light years or approximately $10^{61}$ Planck lengths, which is an enormously larger 4-density.  Therefore, unless the quantum probability of history 1 were greater than the quantum probability of history 2 by this ratio, say $P(h=1|i)/P(h=2|i) > e^{10^{42}}$, history 2 would contribute more to the observational probabilities and give the prediction that most observations would be by ordinary observers rather than by Boltzmann brains, in this 4-volume averaging.

This result implies that unless the quantum probabilities of very long histories are enormously greater than those of histories that last only of the order of the lifetime of the existence of the set of ordinary observers, the observational probabilities will be dominated by the shorter histories if 4-volume averaging is correct.  This would mean that even if one needed a fine tuning of one part in several hundred to set some microphysical parameter (e.g, the gravitino mass or the Higgs mass, as discussed above), if the multiverse included these possibilities and if the quantum probabilities were not minuscule for histories with the property that they would decay on astronomical decay times, then such histories would dominate for observational probabilities, solving the Boltzmann brain problem.

Besides the possibility of producing an astronomically short history by bubble decay, one might also produce it by having the dark energy or quintessence being a scalar field on a nonflat potential that drops below zero, as discussed in \cite{Page05}.  Then, although we may be presently on a part of the potential that is positive, giving the currently observed acceleration of the universe, as the scalar field slowly slides down the very gradual potential slope, the potential would eventually go negative, leading to the universe ending in a big crunch.  For this to be consistent with present observations of the ratio $w$ of the pressure to the energy density of the dark energy, $-1.14 < w << -0.88$ from WMAP, baryon acoustic oscillations, and Type 1a supernovae data for a constant $w$ \cite{WMAP5}, the slope of the potential would have to be extremely small in Planck units, which would apparently require very fine tuning that did not have a previous anthropic explanation.  However, with 4-volume weighting, this fine tuning would be explained anthropically if it were the mechanism to keep the 4-volume small enough that the 4-volume density of observers is not nearly so low as it would be asymptotically from Boltzmann brains for an extremely long lasting universe.

Of course, even if our universe history were most probably astronomically short (not enormously longer than its present age or the asymptotic Hubble time from the present value of the dark energy, after conditioning upon the existence of observations), one would still like a prediction as to what decay mechanism is most likely to produce this.  For example, is it by bubble nucleation, and if so, what microphysical process leads to that?  On the other hand, is it by sliding down the potential for a scalar quintessence, and if so, what is the expected probability distribution for the present value of $w$?  Without having a more detailed prediction of the landscape of possibilities and of their quantum probabilities in the multiverse, it appears difficult to answer such questions.  However, it would be interesting to attempt to answer them, particularly for predicting whether there is likely to be any observational consequence (before the universe actually decays!) of an impending decay.  For example, is it actually likely that we could get a confirmation of the present prediction of decay from the value of the Higgs mass and other microphysical parameters that could be measured at CERN \cite{ADSV,EEGHR}?

\section{Challenges for 4-volume averaging}

4-volume averaging may solve the Boltzmann brain problem by having observational probabilities dominated by histories of the universe (4-geometries with their matter fields) that do not last long enough for significant Boltzmann brain production.  Furthermore, it may do so by a decay mechanism that leads to other observable consequences (e.g., the Higgs mass within a certain range \cite{ADSV,EEGHR} or else a $w > -1$).  It would be very exciting to confirm any of these consequences of the prediction.

However, I should also caution that 4-volume weighting is just one possibility among many others that I have discussed \cite{cmwvw,insuff,brd,BA}, and it does have challenges facing it.  I personally think it is very important to analyze further, but at present I am still rather sceptical that it is right.

My main theoretical scepticism is that normally quantum theory gives amplitudes and expectation values (some of which may be interpreted as probabilities) for configurations or for observables, but not for directly for histories.  There is the program for calculating probabilities for members of sets of decohering histories
\cite{Griffiths,Omnes,GellMann-Hartle,Hartle,Albrecht,Halliwell,Isham, Goldstein-Page,Isham-Linden,Diosi,Dowker-Kent,HLM,Kent,Kent-McElwaine, Hartle-Marolf,Griffiths-Hartle,Brun-Hartle,Markopoulou,Savvidou,Halliwell-Thorwart, Craig-Hartle,Halliwell-Wallden}, but so far it has always struck me as rather {\it ad hoc} which set of decoherent histories to choose for assigning probabilities to members of that set.  So personally I have not been ready to adopt this program myself, but if it leads to a better solution of the Boltzmann brain problem than others and also leads to possibly testable predictions, as using 4-volume weighting within it seems to do, then it would certainly be worth considering.  If so, the main challenge or mystery in my eyes would be what it is that specifies the choice of the set of decohering histories that should be assigned the quantum probabilities.

If one does accept the difficult idea that histories have quantum probabilities (and not just amplitudes in a path integral or sum over histories) and tentatively accepts the idea of 4-volume weighting to convert the quantum probabilities of entire histories to observational probabilities (the probabilities of individual observational results, of which there can be many of varying frequency within each history), there are still an observational challenge facing the present proposal.  This challenge is that if one accepts the anthropic multiverse explanation of the observed cosmic acceleration or effective cosmological constant, with 4-volume weighting it would seem more natural for it to come out as a slightly negative minimum in a potential, so that the universe would evolve into a big crunch directly, without require any bubble formation or slow rolling of a quintessence scalar field down a gentle slope to get to negative values.  

If one assumed that for some unknown reason the present value of the potential had to be positive, one might justify astronomically rapid bubble nucleation or a slight tilt to the potential to cause the universe to decay quickly and make the 4-volume of our history small enough for ordinary observers to dominate over Boltzmann brains.  That is, the enormously extra observational weights in Eq. (\ref{unnormalized-probabilities}) from the $N(j|h)/V_4(h)$ factors would dominate over the greatly reduced quantum probabilities $P(h|i)$ for the histories with the fine tunings for the bubble nucleation rate or for the slight tilt to the quintessence scalar potential.  However, so far as I know, there is no known reason why the observed cosmological constant had to come out positive rather than negative.  If a landscape of possibilities (e.g., from string/M theory) can predict a minimum near zero to be consistent with our observations that it is about 122 orders of magnitude smaller than the Planck density, then there is no reason obvious to me why this minimum should have one sign or the other, that is, why the quantum probabilities for histories should very strongly prefer one sign over the other for values near zero.

If one had a history with a negative minimum for the quintessence scalar potential, the scalar could just sit there without any fine tuning and produce a short-lived universe after the energy density of the other matter fields dropped low enough for the effective negative cosmological constant to dominate and lead to a big crunch.  Therefore, if 4-volume weighting were correct, it seems mysterious why we have not observed a negative cosmological constant.  (If I had been writing this paper before the discovery of the sign of the effective cosmological constant, I would have predicted that 4-volume weighting would most likely lead to an observed effective cosmological constant that is negative.)

I do not presently see how to meet this challenge (or the one of how to specify a particular set of decohering histories so that one can really get probabilities for 4-geometries rather than just for 3-geometries), so the present proposal is certainly incomplete.  Perhaps there is some yet-to-be-discovered reason why having decay by bubble nucleation or by a tilt to the quintessence scalar potential does not have a quantum probability much lower than simply having a slightly negative potential minimum for the quintessence scalar to sit in.

\section{Conclusions}

If quantum theory really can give probabilities for 4-geometries (spacetime histories) rather than just for 3-geometries (spatial configurations), then 4-volume weighting would be a rather natural way of extracting probabilities of observational results, by weighting each history's contribution to the probability of a specific observation not only by the quantum probability of that history but also by the 4-volume density of that observation within the history.  4-volume weighting appears to predict that the universe should not last enormously longer than the lifetime of the set of ordinary observers, so it would give a simple solution of the Boltzmann brain problem.  Depending upon the decay mechanism of our history, 4-volume weighting might lead to predictions that the Higgs mass is in a very narrow range \cite{ADSV,EEGHR} or to predictions that $w > -1$, which would be very interesting to test.

On the other hand, I myself do not understand how to get unique quantum probabilities for 4-geometries rather than for 3-geometries, and I also do not understand how 4-volume weighting should not have led to the prediction that the observed value of the cosmological constant would be negative.  Therefore, the proposal and its predictions are certainly not complete.  However, since the possible consequences of 4-volume averaging are so interesting, I am putting it out for consideration in the hopes that others can help see whether it can meet the challenges facing it and indeed make important predictions for the fate of our universe.

\section*{Acknowledgments}

I am grateful for discussions with Andreas Albrecht, Tom Banks, Raphael
Bousso, Sean Carroll, Brandon Carter, Ben Freivogel, Jaume Garriga, Alan Guth, Daniel Harlow, James Hartle, Thomas Hertog, Gary Horowitz, Matthew Kleban, Andrei Linde, Seth Lloyd, Juan Maldacena, Donald Marolf, Mahdiyar Noorbala, Daniel Phillips, Mark Srednicki, Herman Verlinde, Alex Vilenkin, Alexander Westphal, and others.  I have appreciated the hospitality of the Perimeter Institute for Theoretical Physics, where this idea was written up.  This research was supported in part by the Natural Sciences and Engineering Research Council of Canada.


\newpage

\baselineskip 5pt

\begin{thebibliography}{99}

\bibitem{LLM} A.~Linde, D.~Linde, and A.~Mezhlumian, Phys.\ Rev.\ D {\bf 49},
1783-1826 (1994) [arXiv:gr-qc/9306035].

\bibitem{BLL94} J.~Garcia-Bellido, A.~D.~Linde and D.~A.~Linde, 
Phys.\ Rev.\ D {\bf 50}, 730 (1994) [arXiv:astro-ph/9312039].

\bibitem{Vil95a} A.~Vilenkin, Phys.\ Rev.\ Lett.\ {\bf 74}, 846-849 (1995)
[arXiv:gr-qc/9406010].

\bibitem{GL} J.~Garcia-Bellido and A.~Linde, Phys.\ Rev.\ D {\bf 51}, 429-443
(1995) [arXiv:hep-th/9408023].

\bibitem{LLM2} A.~Linde, D.~Linde, and A.~Mezhlumian, Phys.\ Lett.\ B {\bf 345}, 203-210 (1995) [arXiv:hep-th/9411111].

\bibitem{Vil95b} A.~Vilenkin, Phys.\ Rev.\ D {\bf 52}, 3365-3374 (1995)
[arXiv:gr-qc/9505031].

\bibitem{WV} S.~Winitzki and A.~Vilenkin, Phys.\ Rev.\ D {\bf 53}, 4298-4310
(1996) [arXiv:gr-qc/9510054].

\bibitem{LM} A.~D.~Linde and A.~Mezhlumian, Phys.\ Rev.\ D {\bf 53}, 4267-4274 (1996) [arXiv:gr-qc/9511058].

\bibitem{Page97} D.~N.~Page, Phys.\ Rev.\ D {\bf 56}, 2065-2072 (1997)
[arXiv:gr-qc/9704017].

\bibitem{Vil98} A.~Vilenkin, Phys.\ Rev.\ Lett.\ {\bf 81}, 5501-5504 (1998)
[arXiv:hep-th/9806185].

\bibitem{VVW} V.~Vanchurin, A.~Vilenkin, and S.~Winitzki, Phys.\ Rev.\ D {\bf
61}, 083507 (2000) [arXiv:gr-qc/9905097].

\bibitem{Guth00} A.~H.~Guth, Phys.\ Rept.\ {\bf 333}, 555-574 (2000)
[arXiv:astro-ph/0002156].

\bibitem{GV} J.~Garriga and A.~Vilenkin, Phys.\ Rev.\ D {\bf 64}, 023507 (2001) [arXiv:gr-qc/0102090].

\bibitem{EKS} G.~F.~R.~Ellis, U.~Kirchner, and W.~R.~Stoeger, S.J., Mon.\ Not.\ Roy.\ Astron.\ Soc.\ {\bf 347}, 921-936 (2004) [arXiv:astro-ph/0305292].

\bibitem{SEK} W.~R.~Stoeger, G.~F.~R.~Ellis, and U.~Kirchner, ``Multiverses and Cosmology:  Philosophical Issues,'' arXiv:astro-ph/0407329.

\bibitem{AT} A.~Aguirre and M.~Tegmark, JCAP {\bf 0501}, 003 (2005)
[arXiv:hep-th/0409072].

\bibitem{Teg} M.~Tegmark, JCAP {\bf 0504}, 001 (2005) [arXiv:astro-ph/0410281].

\bibitem{Aguirre} A.~Aguirre, in {\em Universe or Multiverse?}, edited by
B.~J.~Carr (Cambridge University Press, Cambridge, 2007), pp.\ 367-386
[arXiv:astro-ph/0506519].

\bibitem{Ellis} G.~Ellis, in {\em Universe or Multiverse?}, edited by B.~J.~Carr (Cambridge University Press, Cambridge, 2007), pp.\ 387-409.

\bibitem{Page-in-Carr} D.~N.~Page, in {\it Universe or Multiverse?}, edited by B.~J.~Carr (Cambridge University Press, Cambridge, 2007), pp.\ 411-429
[arXiv:hep-th/0610101]. 

\bibitem{GSVW} J.~Garriga, D.~Schwartz-Perlov, A.~Vilenkin, and S.~Winitzki,
JCAP {\bf 0601}, 017 (2006) [arXiv:hep-th/0509184].

\bibitem{ELM} R.~Easther, E.~A.~Lim, and M.~R.~Martin, JCAP {\bf 0603}, 016
(2006) [arXiv:astro-ph/0511233].

\bibitem{BFL} R.~Bousso, B.~Freivogel, and M.~Lippert, Phys.\ Rev.\ D {\bf 74}, 046008 (2006) [arXiv:hep-th/0603105].

\bibitem{Bou06} R.~Bousso, Phys.\ Rev.\ Lett.\ {\bf 97}, 191302 (2006)
[arXiv:hep-th/0605263].

\bibitem{CDGKL} A.~Ceresole, G.~Dall'Agata, A.~Giryavets, R.~Kallosh, and
A.~Linde, Phys.\ Rev.\ D {\bf 74}, 086010 (2006) [arXiv:hep-th/0605266].

\bibitem{BFY06} R.~Bousso, B.~Freivogel, and I-S.\ Yang, Phys.\ Rev.\ D {\bf
74}, 103516 (2006) [arXiv:hep-th/0606114].

\bibitem{GT} G.~W.~Gibbons and N.~Turok, Phys.\ Rev.\ D {\bf 77}, 063516 (2008) [arXiv:hep-th/0609095].

\bibitem{Vil07} A.~Vilenkin, J.\ Phys.\ A {\bf 40}, 6777-6785 (2007)
[arXiv:hep-th/0609193].

\bibitem{AGJ1} A.~Aguirre, S.~Gratton, and M.~C.~Johnson, Phys.\ Rev.\ D {\bf
75}, 123501 (2007) [arXiv:hep-th/0611221].

\bibitem{Win06} S.~Winitzki, Lect.\ Notes Phys.\ {\bf 738}, 157 (2008)
[arXiv:gr-qc/0612164].

\bibitem{AGJ2} A.~Aguirre, S.~Gratton, and M.~C.~Johnson, Phys.\ Rev.\ Lett.\
{\bf 98}, 131301 (2007) [arXiv:hep-th/0612195].

\bibitem{BHKP} R.~Bousso, R.~Harnik, G.~D.~Kribs, and G.~Perez, Phys.\ Rev.\ D {\bf 76}, 043513 (2007) [arXiv:hep-th/0702115].

\bibitem{Guth07} A.~H.~Guth, J.\ Phys.\ A {\bf 40}, 6811-6826 (2007)
[arXiv:hep-th/0702178].

\bibitem{BY} R.~Bousso and I-S.\ Yang, Phys.\ Rev.\ D {\bf 75} 123520 (2007)
[arXiv:hep-th/0703206].

\bibitem{LWa} M.~Li and Y.~Wang, JCAP {\bf 0706}, 012 (2007) [arXiv:0704.1026
[hep-th]].

\bibitem{ADNTV} N.~Arkani-Hamed, S.~Dubovsky, A.~Nicolis, E.~Trincherini, and
 G.~Villadoro, JHEP {\bf 0705}, 055 (2007) [arXiv:0704.1814 [hep-th]].

\bibitem{Linde08a} A.~Linde, Lect.\ Notes Phys.\ {\bf 738}, 1-54 (2008)
[arXiv:0705.0164 [hep-th]].

\bibitem{Linde08b} A.~Linde, JCAP {\bf 0706}, 017 (2007) [arXiv:0705.1160 [hep-th]].

\bibitem{LWb} M.~Li and Y.~Wang, JCAP {\bf 0708}, 007 (2007) [arXiv:0706.1691
[hep-th]].

\bibitem{CSS} T.~Clifton, S.~Shenker, and N.~Sivanandam, JHEP {\bf 0709}, 034 (2007) [arXiv:0706.3201 [hep-th]].

\bibitem{Haw} S.~W.~Hawking, ``Volume Weighting in the No Boundary Proposal,'' arXiv:0710.2029 [hep-th].

\bibitem{GV07} J.~Garriga and A.~Vilenkin, Phys.\ Rev.\ D {\bf 77}, 043526
(2008) [arXiv:0711.2559 [hep-th]].

\bibitem{HHH} J.~B.~Hartle, S.~W.~Hawking, and T.~Hertog, Phys.\ Rev.\ Lett.\
{\bf 100}, 201301 (2008) [arXiv:0711.4630 [hep-th]]; Phys.\ Rev.\ D {\bf 77},
123537 (2008) [arXiv:0803.1663 [hep-th]].

\bibitem{Win08a} S.~Winitzki, Phys.\ Rev.\ D {\bf 78}, 043501 (2008)
[arXiv:0803.1300 [gr-qc]].

\bibitem{SGSV} A.~De Simone, A.~H.~Guth, M.~P.~Salem, and A.~Vilenkin,
Phys.\ Rev.\ D {\bf 78}, 063520 (2008) [arXiv:0805.2173 [hep-th]].

\bibitem{BFYb} R.~Bousso, B.~Freivogel, and I-S.\ Yang, Phys.\ Rev.\ D {\bf 79}, 063513 (2009), [arXiv:0808.3770 [hep-th]].

\bibitem{DGLNSV} A.~De Simone, A.~H.~Guth, A.~Linde, M.~Noorbala, M.~P.~Salem, and A.~Vilenkin, ``Boltzmann Brains and the Scale-Factor Cutoff Measure of the Multiverse,'' arXiv:0808.3778 [hep-th].

\bibitem{GV09} J.~Garriga and A.~Vilenkin, JCAP {\bf 0901}, 021 (2009) [arXiv:0809.4257 [hep-th]].

\bibitem{Win08b} S.~Winitzki, Phys.\ Rev.\ D {\bf 78}, 063517 (2008) [arXiv:0805.3940 [gr-qc]].

\bibitem{Wang} Y.~Wang, ``Eternal Inflation: Prohibited by Quantum Gravity?'' arXiv:0805.4520 [hep-th].

\bibitem{Win08c} S.~Winitzki, Phys.\ Rev.\ D {\bf 78}, 123518 (2008) [arXiv:0810.1517 [gr-qc]].

\bibitem{McI} B.~McInnes, ``Horizon Complementarity and Casimir Violations of the Null Energy Condition,'' arXiv:0811.4465 [hep-th].

\bibitem{LVW} A.~Linde, V.~Vanchurin, and S.~Winitzki, JCAP {\bf 0901}, 031 (2009) [arXiv:0812.0005 [hep-th]].

\bibitem{HaHe} J.~B.~Hartle and T.~Hertog, ``Replication Regulates Volume Weighting in Quantum Cosmology,'' arXiv:0905.3877 [hep-th].

\bibitem{SH} M.~Srednicki and J.~B.~Hartle, ``Science in a Very Large Universe,'' arXiv:0906.0042.

\bibitem{Albrecht09} A.~Albrecht, ``De Sitter Equilibrium as a Fundamental Framework for Cosmology,'' arXiv:0906.1047 [gr-qc].

\bibitem{DKS} L.~Dyson, M.~Kleban, and L.~Susskind, JHEP {\bf 0210}, 011 (2002) [arXiv:hep-th/0208013].

\bibitem{Albrecht04} A.~Albrecht, in {\em Science and Ultimate Reality:  Quantum Theory, Cosmology, and Complexity}, edited by J.~D.~Barrow, P.~C.~W.~Davies, and C.~L.~Harper, Jr.\ (Cambridge University Press, Cambridge, 2004), pp. 363-401 [arXiv:astro-ph/0210527].

\bibitem{AS} A.~Albrecht and L.~Sorbo, Phys.\ Rev.\ D {\bf 70}, 063528 (2004) [arXiv:hep-th/0405270].

\bibitem{Page05} D.~N.~Page, J.\ Korean Phys.\ Soc.\ {\bf 49}, 711-714 (2006) [arXiv:hep-th/0510003].

\bibitem{YY} A.~V.~Yurov and V.~A.~Yurov, ``One More Observational Consequence of Many-Worlds Quantum Theory,'' arXiv:hep-th/0511238.

\bibitem{Page06a} D.~N.~Page,  Phys.\ Rev.\ D {\bf 78}, 063535 (2008) [arXiv:hep-th/0610079].

\bibitem{BF} R.~Bousso and B.~Freivogel, J.\ High Energy Phys.\ {\bf 0706},
018 (2007) [arXiv:hep-th/0610132]. 

\bibitem{Page06b} D.~N.~Page, J.\ Cosmolog.\ Astropart.\ Phys.\ {\bf 0701},
004 (2007) [arXiv:hep-th/0610199].

\bibitem{Bou05} R.~Bousso, ``Precision Cosmology and the Landscape,'' arXiv:hep-th/0610211.

\bibitem{Linde06} A.~Linde, J.\ Cosmolog.\ Astropart.\ Phys.\ {\bf 0701},
022 (2007) [arXiv:hep-th/0611043]. 

\bibitem{Page06c} D.~N.~Page, Phys.\ Rev.\ D {\bf 78}, 063536 (2008) [arXiv:hep-th/0611158]. 

\bibitem{Vil06}A.~Vilenkin, J.\ High Energy Phys.\ {\bf 0701}, 092 (2007) [arXiv:hep-th/0611271].

\bibitem{Page06d} D.~N.~Page, Phys.\ Lett.\ B {\bf 669}, 197-200 (2008) [arXiv:hep-th/0612137].

\bibitem{Vanchurin} V.~Vanchurin, Phys.\ Rev.\ D {\bf 75}, 023524 (2007)
[arXiv:hep-th/0612215].

\bibitem{Banks} T.~Banks, ``Entropy and Initial Conditions in Cosmology,''
arXiv:hep-th/0701146.

\bibitem{Carlip} S.~Carlip, J.\ Cosmolog.\ Astropart.\ Phys.\ {\bf 0706},
001 (2007) [arXiv:hep-th/0703115].

\bibitem{HS} J.~B.~Hartle and M.~Srednicki, Phys.\ Rev.\ D {\bf 75}, 123523
(2007) [arXiv:0704.2630].

\bibitem{GM} S.~B.~Giddings and D.~Marolf, Phys.\ Rev.\ D {\bf 76}, 064023
(2007) [arXiv:0705.1178 [hep-th]].

\bibitem{Giddings} S.~B.~Giddings, Mod.\ Phys.\ Lett.\ A {\bf 22}, 2949-2954
(2007) [arXiv:arXiv:0705.2197 [hep-th]].

\bibitem{typdef} D.~N.~Page, ``Typicality Defended,'' arXiv:0707.4169 [hep-th].  

\bibitem{LWc} M.~Li and Y.~Wang, ``Typicality, Freak Observers and the Anthropic Principle of Existence,'' arXiv:0708.4077 [hep-th].

\bibitem{DP07} D.~N.~Page, in
{\em Proceedings of 13th International Congress of Logic,
Methodology and Philosophy of Science}, edited by C.\ Glymour, W.\ Wang, and
D.\ Westerst\aa hl (Kings College Publications, 2008) [arXiv:0712.2240 [hep-th]].

\bibitem{Bou08} R.~Bousso, Gen.\ Rel.\ Grav.\ {\bf 40}, 607-637 (2008) [arXiv:0708.4231 [hep-th]]. 

\bibitem{BFYa} R.~Bousso, B.~Freivogel, and I-S.\ Yang, Phys.\ Rev.\ D {\bf 77}, 103514 (2008) [arXiv:0712.3324 [hep-th]].

\bibitem{Gott} J.~R.~Gott, III, ``Boltzmann Brains: I'd Rather See than Be
One,'' arXiv:0802.0233 [gr-qc].

\bibitem{typder} D.~N.~Page, Phys.\ Rev.\ D {\bf 78}, 023514 (2008) [arXiv:0804.3592 [hep-th]].

\bibitem{FL} B.~Freivogel and M.~Lippert, JHEP {\bf 0812}, 096 (2008) [arXiv:0807.1104 [hep-th]].

\bibitem{ADSV} N.~Arkani-Hamed, S.~Dubovsky, L.~Senatore, and G.~Villadoro,
J.\ High Energy Phys.\ {\bf 0803}, 075 (2008) [arXiv:0801.2399 [hep-ph]].

\bibitem{cmwvw} D.~N.~Page, J.\ Cosmolog.\ Astropart.\ Phys.\ {\bf
0810}, 025 (2008), arXiv:0808.0351 [hep-th].

\bibitem{insuff} D.~N.~Page, Phys.\ Lett.\ B {\bf 678}, 41-44 (2009) [arXiv:0808.0722 [hep-th]].

\bibitem{brd} D.~N.~Page, J.\ Cosmolog.\ Astropart.\ Phys.\ {\bf 0708},
008 (2009) [arXiv:0903.4888 [hep-th]].

\bibitem{BA} D.~N.~Page, ``Born Again,'' arxiv:0907.4152 [hep-th].

\bibitem{Smolin} L.~Smolin, ``The Status of Cosmological Natural Selection,'' arXiv:hep-th/0612185.

\bibitem{BS} J.~D.~Barrow and D.~J.~Shaw, Class.\ Quant.\ Grav.\ {\bf 25}, 085012 (2008) [arXiv:0712.2190 [gr-qc]].

\bibitem{MA} L.~Mersini-Houghton and F.~C.~Adams, Class.\ Quant.\ Grav.\ {\bf 25}, 165002 (2008) [arXiv:0810.4914 [gr-qc]].

\bibitem{EEGHR} J.~Ellis, J.~R.~Espinosa, G.~F.~Giudice, A.~Hoecker, and A.~Riotto, ``The Probable Fate of the Standard Model,'' arXiv:0906.0954 [hep-ph].

\bibitem{Vet} A.~Vilenkin, Phys.\ Rev.\ {\bf D27}, 2848-2855 (1983); Nucl.\
Phys.\ Proc.\ Suppl.\ {\bf 88}, 67-74 (2000) [arXiv:gr-qc/9911087]; ``Eternal
Inflation and Chaotic Terminology,'' arXiv:gr-qc/0409055; {\em Many Worlds in
One: The Search for Other Universes} (Hill and Wang, New York, 2006).

\bibitem{Let} A.~D.~Linde, in {\em  The Very Early Universe}, ed.\
G.~W.~Gibbons, S.~W.~Hawking and S.~Siklos, Cambridge University Press (1983), pp.\ 205-249 [http://www.stanford.edu/$\sim$alinde/1983.pdf]; Mod.\ Phys.\ Lett.\ {\bf A1}, 81-85 (1986); Phys.\ Lett.\ {\bf B175}, 395-400 (1986); Phys.\ Scripta {\bf T15}, 169 (1987); Phys.\ Lett.\ {\bf B249}, 18-26 (1990);  {\it Particle  Physics  and Inflationary Cosmology} (Harwood, Chur, Switzerland, 1990) [arXiv:hep-th/0503203]; Sci.\ Am.\ {\bf 271}, 32-39 (1994).

\bibitem{Set} A.~A.~Starobinsky, in {\em Field Theory, Quantum
Gravity, and Strings}, edited by H.~J.~de Vega and N.~Sanchez,
Lecture Notes in Physics Vol.\ 246 (Springer, Heidelberg, 1986).

\bibitem{Get} A.~H.~Guth, Phys.\ Rept.\ {\bf 333}, 555-574 (2000)
[arXiv:astro-ph/0002156].

\bibitem{HP} S.~W.~Hawking and D.~N.~Page, Nucl.\ Phys.\ B {\bf 264}, 185-196
(1986).

\bibitem{Griffiths} R.~B.~Griffiths,
  J.\ Statist.\ Phys.\  {\bf 36}, 219 (1984);
  ``Consistent Histories and Quantum Reasoning,
  arXiv:quant-ph/9606004;
  Fortsch.\ Phys.\  {\bf 46}, 741 (1998)
  [arXiv:quant-ph/9810016];
  Phys.\ Lett.\  A {\bf 261}, 227 (1999)
  [arXiv:quant-ph/9902059].

\bibitem{Omnes}
  R.~Omnes,
  ``Proposal for an Objective Formulation of Quantum Mechanics,''
  LPTHE-ORSAY-87/07 (1987);
  ``Reformulating the Experimental Interpretation of Quantum Mechanics,''
  LPTHE-ORSAY-87/28 (1987);
  ``Interpretation of Quantum Mechanics,''
  LPTHE-ORSAY-87/34 (1987);
  ``How Mister Tompkins Understood Quantum Mechanics,''
  LPTHE-ORSAY-87/65 (1987);
  J.\ Statist.\ Phys.\  {\bf 53}, 893 (1988);
  J.\ Statist.\ Phys.\  {\bf 53}, 933 (1988);
  J.\ Statist.\ Phys.\  {\bf 53}, 957 (1988);
  J.\ Statist.\ Phys.\  {\bf 57}, 356 (1989);
  ``About the Notion of Truth in Quantum Mechanics,''
  PRINT-90-0462-ORSAY- (1990);
  ``General Theory of the Decoherence Effect in Quantum Mechanics,''
  LPTHE-ORSAY-92-102 (1992);
  Rev.\ Mod.\ Phys.\  {\bf 64}, 339 (1992);
  ``A New Interpretation of Quantum Mechanics and its Consequence in
  Epistemology,''
  LPTHE-ORSAY-94-32 (1994);
  ``New Interpretations of Quantum Mechanics and the Theory of Knowledge,''
  LPTHE-ORSAY-94-86 (1994);
  Phys.\ Rev.\  D {\bf 71}, 065011 (2005)
  [arXiv:quant-ph/0411201].

\bibitem{GellMann-Hartle}
  M.~Gell-Mann and J.~B.~Hartle,
  ``Quantum Mechanics in the Light of Quantum Cosmology,''
  PRINT-90-0266-UC,SANTA-BARBARA- (1990);
  ``Alternative Decohering Histories in Quantum Mechanics,''
  UCSB-TH-90-56 (1990);
  ``Time Symmetry and Asymmetry in Quantum Mechanics and Quantum Cosmology,''
  arXiv:gr-qc/9304023;
  Phys.\ Rev.\  D {\bf 47}, 3345 (1993)
  [arXiv:gr-qc/9210010];
  ``Equivalent Sets of Histories and Multiple Quasiclassical Domains,''
  arXiv:gr-qc/9404013;
  ``Strong Decoherence,''
  arXiv:gr-qc/9509054;
  Phys.\ Rev.\  A {\bf 76}, 022104 (2007)
  [arXiv:quant-ph/0609190].

\bibitem{Hartle}
  J.~B.~Hartle,
  Vistas Astron.\  {\bf 37}, 569 (1993)
  [arXiv:gr-qc/9210004];
  ``Space-Time Quantum Mechanics and the Quantum Mechanics of Space-Time,''
  arXiv:gr-qc/9304006;
  ``The Quantum Mechanics of Closed Systems,''
  arXiv:gr-qc/9210006;
  ``The Reduction of the State vector and Limitations on Measurement in the
  Quantum Mechanics of Closed Systems,''
  arXiv:gr-qc/9301011;
  Phys.\ Rev.\  D {\bf 49}, 6543 (1994)
  [arXiv:gr-qc/9309012];
  ``Quasiclassical Domains in a Quantum Universe,''
  arXiv:gr-qc/9404017;
  Phys.\ Rev.\  D {\bf 51}, 1800 (1995)
  [arXiv:gr-qc/9409005];
  ``Quantum Mechanics at the Planck Scale,''
  arXiv:gr-qc/9508023;
  Phys.\ Scripta {\bf T76}, 67 (1998)
  [arXiv:gr-qc/9712001];
  ``Generalized Quantum Theory and Black Hole Evaporation,''
  arXiv:gr-qc/9808070;
  Int.\ J.\ Mod.\ Phys.\  A {\bf 16} 1 (2001);
  Phys.\ Rev.\  A {\bf 69}, 042111 (2004)
  [arXiv:quant-ph/0209104];
  arXiv:quant-ph/0305089;
  ``Linear Positivity and Virtual Probability,''
  arXiv:quant-ph/0401108;
  Int.\ J.\ Theor.\ Phys.\  {\bf 45}, 1390 (2006)
  [arXiv:gr-qc/0510126];
  J.~B.~Hartle,
  ``Generalizing Quantum Mechanics for Quantum Spacetime,''
  arXiv:gr-qc/0602013;
  ``Quantum Mechanics with Extended Probabilities,''
  arXiv:0801.0688 [quant-ph];
  ``The Quasiclassical Realms of this Quantum Universe,''
  arXiv:0806.3776 [quant-ph];

\bibitem{Albrecht}
  A.~J.~Albrecht,
  Phys.\ Rev.\  D {\bf 48}, 3768 (1993)
  [arXiv:hep-th/9309051].

\bibitem{Halliwell}
  J.~J.~Halliwell,
  ``Aspects of the Decoherent Histories Approach to Quantum Mechanics,''
  arXiv:gr-qc/9308005;
  Annals N.\ Y.\ Acad.\ Sci.\  {\bf 755}, 726 (1995)
  [arXiv:gr-qc/9407040];
  Phys.\ Rev.\  D {\bf 58}, 105015 (1998)
  [arXiv:quant-ph/9805062];
  Phys.\ Rev.\  D {\bf 60}, 105031 (1999)
  [arXiv:quant-ph/9902008];
  Phys.\ Rev.\  D {\bf 63}, 085013 (2001)
  [arXiv:quant-ph/0011103];
  ``Decoherent Histories for Spacetime Domains,''
  arXiv:quant-ph/0101099.

\bibitem{Isham}
  C.~J.~Isham,
  J.\ Math.\ Phys.\  {\bf 35}, 2157 (1994)
  [arXiv:gr-qc/9308006];
  Int.\ J.\ Theor.\ Phys.\  {\bf 36}, 785 (1997)
  [arXiv:gr-qc/9607069].

\bibitem{Goldstein-Page}
  S.~Goldstein and D.~N.~Page,
  Phys.\ Rev.\ Lett.\  {\bf 74}, 3715 (1995)
  [arXiv:gr-qc/9403055].

\bibitem{Isham-Linden}
  C.~J.~Isham and N.~Linden,
  J.\ Math.\ Phys.\  {\bf 35}, 5452 (1994)
  [arXiv:gr-qc/9405029];
  J.\ Math.\ Phys.\  {\bf 36}, 5392 (1995)
  [arXiv:gr-qc/9503063].

\bibitem{Diosi}
  L.~Diosi,
  Phys.\ Lett.\  {\bf 203A}, 267 (1995)
  [arXiv:gr-qc/9409028];
  Phys.\ Rev.\ Lett.\  {\bf 92}, 170401 (2004)
  [arXiv:quant-ph/0310181].

\bibitem{Dowker-Kent}
  F.~Dowker and A.~Kent,
  Phys.\ Rev.\ Lett.\  {\bf 75}, 3038 (1995)
  [arXiv:gr-qc/9409037];
  J.\ Statist.\ Phys.\  {\bf 82}, 1575 (1996)
  [arXiv:gr-qc/9412067].

\bibitem{HLM}
  J.~B.~Hartle, R.~Laflamme and D.~Marolf,
  Phys.\ Rev.\  D {\bf 51}, 7007 (1995)
  [arXiv:gr-qc/9410006].

\bibitem{Kent}
  A.~Kent,
  Phys.\ Rev.\  A {\bf 54}, 4670 (1996)
  [arXiv:gr-qc/9512023];
  Phys.\ Rev.\ Lett.\  {\bf 78}, 2874 (1997)
  [arXiv:gr-qc/9604012];
  Lect.\ Notes Phys.\  {\bf 559}, 93 (2000)
  [arXiv:gr-qc/9607073];
  Phys.\ Rev.\  D {\bf 56}, 2469 (1997)
  [arXiv:gr-qc/9610075];
  Phys.\ Scripta {\bf T76}, 78 (1998)
  [arXiv:gr-qc/9809026];
  Phys.\ Rev.\ Lett.\  {\bf 81}, 1982 (1998)
  [arXiv:gr-qc/9808016].

\bibitem{Kent-McElwaine}
  A.~Kent and J.~McElwaine,
  Phys.\ Rev.\  A {\bf 55}, 1703 (1997)
  [arXiv:gr-qc/9610028].

\bibitem{Hartle-Marolf}
  J.~B.~Hartle and D.~Marolf,
  Phys.\ Rev.\  D {\bf 56}, 6247 (1997)
  [arXiv:gr-qc/9703021].

\bibitem{Griffiths-Hartle}
  R.~B.~Griffiths and J.~B.~Hartle,
  Phys.\ Rev.\ Lett.\  {\bf 81}, 1981 (1998)
  [arXiv:gr-qc/9710025].

\bibitem{Brun-Hartle}
  T.~A.~Brun and J.~B.~Hartle,
  Phys.\ Rev.\  E {\bf 59}, 6370 (1999)
  [arXiv:quant-ph/9808024].

\bibitem{Markopoulou}
  F.~Markopoulou,
  Class.\ Quant.\ Grav.\  {\bf 17}, 2059 (2000)
  [arXiv:hep-th/9904009];
  Nucl.\ Phys.\ Proc.\ Suppl.\  {\bf 88}, 308 (2000)
  [arXiv:hep-th/9912137].

\bibitem{Savvidou}
  K.~Savvidou,
  ``Continuous Time and Consistent Histories,''
  arXiv:gr-qc/9912076;
  J.\ Math.\ Phys.\  {\bf 43}, 3053 (2002)
  [arXiv:gr-qc/0104053];
  Class.\ Quant.\ Grav.\  {\bf 18}, 3611 (2001)
  [arXiv:gr-qc/0104081];
  Class.\ Quant.\ Grav.\  {\bf 21}, 615 (2004)
  [arXiv:gr-qc/0306034];
  Class.\ Quant.\ Grav.\  {\bf 21}, 631 (2004)
  [arXiv:gr-qc/0306036];
  Braz.\ J.\ Phys.\  {\bf 35}, 307 (2005)
  [arXiv:gr-qc/0412059].

\bibitem{Halliwell-Thorwart}
  J.~J.~Halliwell and J.~Thorwart,
  Phys.\ Rev.\  D {\bf 64}, 124018 (2001)
  [arXiv:gr-qc/0106095];
  Phys.\ Rev.\  D {\bf 65}, 104009 (2002)
  [arXiv:gr-qc/0201070].

\bibitem{Craig-Hartle}
  D.~Craig and J.~B.~Hartle,
  Phys.\ Rev.\  D {\bf 69}, 123525 (2004)
  [arXiv:gr-qc/0309117].

\bibitem{Halliwell-Wallden}
  J.~J.~Halliwell and P.~Wallden,
  Phys.\ Rev.\  D {\bf 73}, 024011 (2006)
  [arXiv:gr-qc/0509013].

\bibitem{DP2006} D.~N.~Page, J.\ Cosmolog.\ Astropart.\ Phys.\ {\bf
0701}, 004 (2007) [arXiv:hep-th/0610199].

\bibitem{WMAP5} E.~Komatsu {\it et al.} (WMAP Collaboration), Astrophys.\ J.\ Suppl.\ {\bf 180}, 330-376 (2009) [arXiv:0803.0547 [astro-ph]].

\end{thebibliography}
\end{document}